\documentclass[a4paper]{jpconf}
\usepackage{graphicx}
\usepackage{hyperref}
\usepackage{amsmath,amssymb}
\usepackage{multirow}
\usepackage{array}
\usepackage[noadjust]{cite}
\usepackage[bottom]{footmisc}

\begin{document}

\title{OAM multiple transmission using uniform circular arrays: numerical modeling and experimental verification with two digital television signals}

\author{Rossella Gaffoglio$^1$, Andrea Cagliero$^1$, Assunta De Vita$^2$ and Bruno Sacco$^2$}

\address{$^1$ Department of Physics, University of Torino, I-10125 Torino, Italy}
\address{$^2$ Centre for Research and Technological Innovation, RAI Radiotelevisione Italiana, I-10128 Torino, Italy}

\address{\textbf{This paper has been accepted for publication by AGU Radio Science, DOI: 10.1002/2015RS005862. It is subject to the American Geophysical Union Copyright. The copy of record is available at AGU Digital Library: \url{http://onlinelibrary.wiley.com/doi/10.1002/2015RS005862/full}}}

\ead{rossella.gaffoglio@unito.it}

\begin{abstract}
In this work we present the outcomes of a radio-frequency OAM transmission between two antenna arrays performed in a real-world context. The analysis is supplemented by deep simulative investigations able to provide both a preliminary overview of the experimental scenario and {\it a posteriori} validation of the achieved results.
As a first step, the far-field OAM communication link is tested at various frequencies and the corresponding link budget is studied by means of an angular scan generated by the rotation of the receiving system. Then, on the same site, two digital television signals encoded as OAM modes ($\ell=1$ and $\ell=-1$) are simultaneously transmitted at a common frequency of $198.5$ MHz with good mode insulation.
\end{abstract}

\section{Introduction} \label{sec_intro}

The last few years have witnessed a growing interest in electromagnetic radiation carrying Orbital Angular Momentum (OAM), which has found relevant applications in  nanotechnologies \cite{Grier}, astronomy \cite{Harwit, Anzolin, Berkhout, Elias}, quantum physics \cite{Molina, Vaziri, Jack}, optics \cite{Gibson, Wang} and radio communications \cite{Thide, Tamburini}. 

OAM beams represent classes of solutions to the Helmholtz equation characterized by a wavefront which consists of $\ell$ intertwined helices, described by the azimuthal phase term $e^{i \ell \varphi}$,  where $\ell$ is the OAM mode index, and a doughnut-shaped beam profile for $\ell \ne 0$. Each class comprises a theoretically infinite set of orthogonal solutions which offers, in principle, the possibility to encode separate channels on the same frequency, an idea that was originally proposed for optical communications \cite{Gibson}. More recently, this aspect and its implications have been deeply explored both theoretically \cite{Thide} and experimentally \cite{Tamburini} also in the RF domain, leading to significant results \cite{Yan, Mari}.

A great amount of possibilities exists to generate OAM beams at various frequency ranges \cite{Padgett, Berry, Salo, Trinder}; among these, one of the most suitable for radio transmissions consists in the use of antenna arrays, where the sought-for solutions can be implemented by means of a proper array synthesis technique \cite{Mohammadi}. 

In the present work we perform the experimental analysis of an OAM-based radio transmission with antenna arrays in the VHF TV band ($174\div230$ MHz) supported by a preliminary numerical study. The experiment represents, to the best of our knowledge, the first case of multiple OAM transmission/reception of Digital Video Broadcasting-Terrestrial (DVB-T) signals using a single array of identical radiators on both sides of the communication link, in the far-field region. One of the main results of the experiment lies in the verification of the feasibility of such transmissions in a real-world context. Moreover, a validation of the ``OAM-link pattern'' concept,  introduced in \cite{Cagliero}, is provided.

Section \ref{sec_Motiv} concerns the description of the experimental set-up and the motivations behind our choices. Then, section \ref{sec_OAMpattern} resumes the fundamental concept of OAM-link pattern, which enables a preliminary estimation of the results expected from measures. Some brief notes on the numerical simulations carried out in order to support the experimental investigation are provided in section \ref{sec_simul}. The technical details on the electronic devices employed for the OAM multiplexing/demultiplexing, the tests set-up and the experimental analysis are explained in section \ref{sec_OAMtrans}. Finally, sections \ref{sec_experiment} and \ref{sec_expRes} report the study and the results relative to the simultaneous transmission of two DVB-T signals encoded as OAM $\ell=1$ and $\ell=-1$ modes, performed with the same experimental apparatus described in the previous part of the work.

\begin{figure}[b]
\centering
\includegraphics[trim=0 0 0 0,clip,width=6.2in]{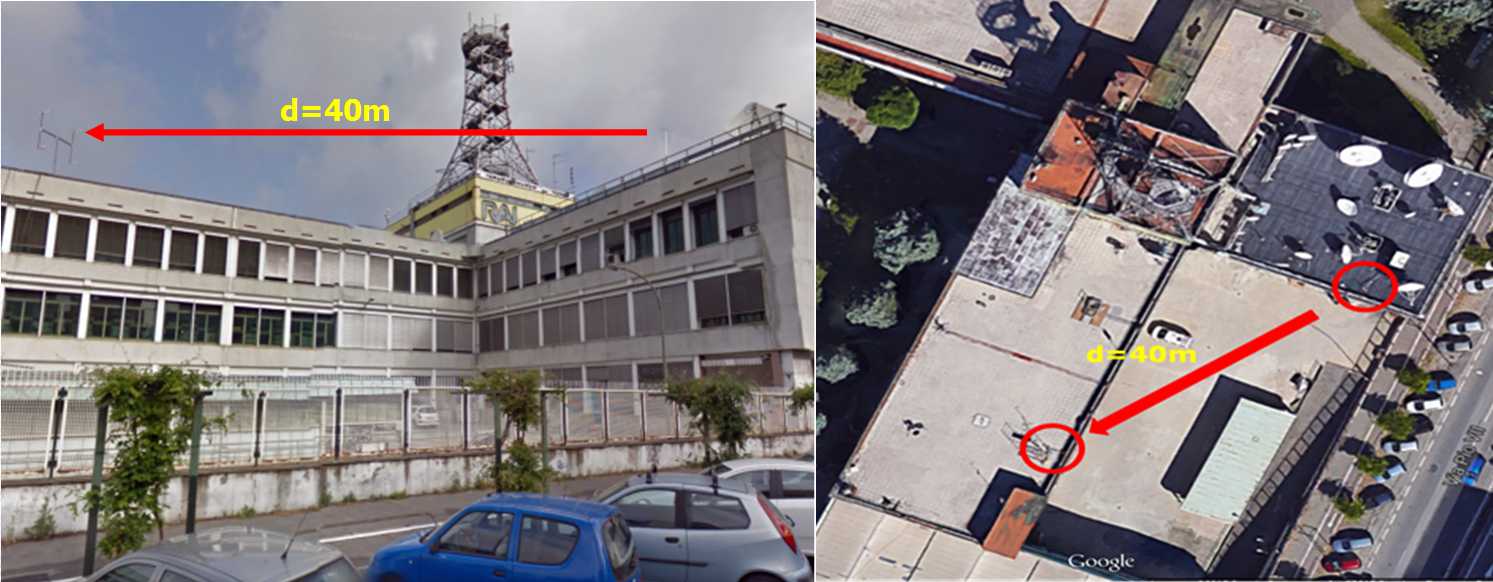}
\caption{Two different views of the experimental site (reproduced from Google Earth, \copyright $\,$2015 Google).}
\label{fig_gaffo1}
\end{figure}

\section{A brief description of the experimental set-up} \label{sec_Motiv}

The motivation behind the choice of working with antenna arrays is dictated by the fact that, although extensively studied in literature both theoretically and numerically \cite{Thide, Mohammadi}, to the best of our knowledge there are no experimental reports of their usage in OAM far-field radio transmissions. Moreover, antenna arrays are better suited to implement a multimode transmission with respect to other OAM antennas. Indeed, a single array fed via beamforming network can support the simultaneous transmission of several independent OAM modes, in contrast to other antennas like the twisted parabolic reflector \cite{Trinder}. In addition, this solution offers the possibility to validate the functionality of specially designed electronic devices conceived as beamforming networks for the OAM mutiplexing and demultiplexing in antenna arrays, designed and patented before the experimental analysis \cite{modesorter}. In the following, we will refer to such devices as OAM ``mode sorters''.

In order to ensure technical simplicity, a minimal number of array elements was considered, allowing to check the OAM concept far from the cases treated in literature, where many more elements are usually taken into account.

As shown in Fig. \ref{fig_gaffo1}, the two systems were placed on the two opposite ends of the flat roof of the RAI Research Centre. The choice of the location for the experiment was essentially motivated by two factors: the height and the arrangement of the building, which permit negligible ground reflection, and the intention to show a true far-field transmission (a link distance of $40$ m corresponds to more than three times the far-field threshold for the case under consideration). 

The decision of working in the frequency band around $205$ MHz, which corresponds to a wavelength $\lambda=1.46$ m, was dictated by the purpose of a TV transmission and by the need to work in a frequency range where the functionalities of the electronic devices were ensured.

\section{An overview of the ``OAM-link pattern'' concept} \label{sec_OAMpattern}

\begin{figure}[t]
\centering
\includegraphics[scale=0.53]{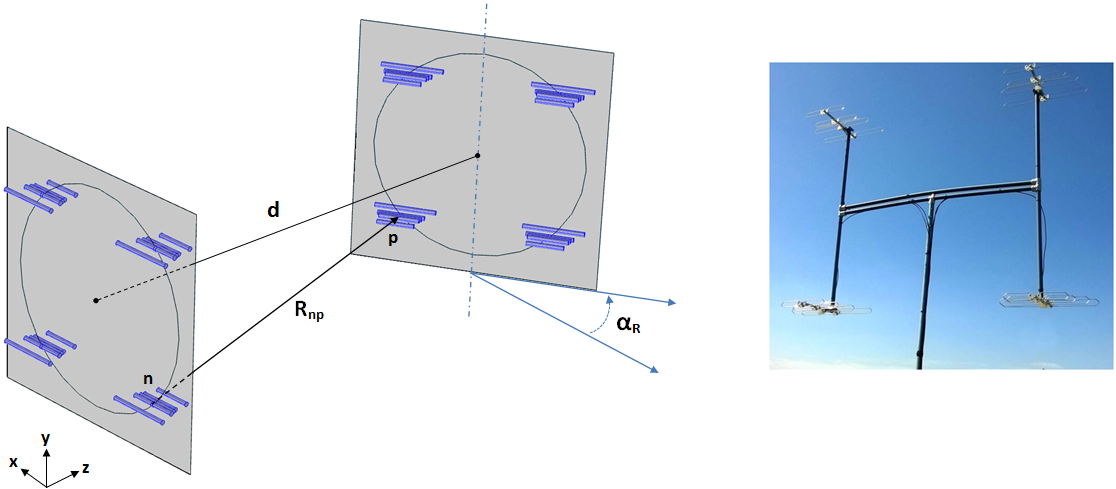}
\caption{Schematic illustration of the link ({\it left}) and a photograph of the transmitting UCA employed in the experiment ({\it right}).}
\label{fig_gaffo2}
\end{figure}

When considering OAM transmissions between antenna arrays, a proper phase-weighting must be introduced at the receiver in order to account for the azimuthal phase profile of a vortex wave and lead to a non-null on-axis power transfer response \cite{Nguyen}. To this end, in \cite{Cagliero}, the following formula for the OAM-link budget estimation between two facing arrays was obtained:
\begin{equation} \label{eq_link2}
\frac{P^{\ell_{T,R}}_{out}}{P_{in}}=\left|\frac{ik\eta}{2R}\frac{1}{\sqrt{N_TN_R}}\sum_{p=1}^{N_R}\sum_{n=1}^{N_T}\frac{e^{-ik r_{np}}}{4\pi r_{np}} \Phi^n_{\ell_T}\Phi_{-\ell_R}^p\mathbf{h}^{\substack{\text{{\tiny{T}}}}}_{np}\cdot\mathbf{h}^{\substack{\text{\tiny{R}}}}_{pn}\right|^2,
\end{equation}
where $k=2\pi/\lambda$ is the modulus of the wave vector, $\eta$ is the vacuum impedance, $R$ is the resistance of the radiators, $N_T$ and $N_R$ indicate the number of elements in the transmitting and the receiving array, respectively, while $\Phi^n_\ell$ is the feed coefficient associated to the $n$th antenna, resulting from the synthesis of an OAM mode with index $\ell$ over the considered array layout. Furthermore, $\mathbf{h}^{\substack{\text{\tiny{T}}}}$ and $\mathbf{h}^{\substack{\text{\tiny{R}}}}$ are the effective heights for each link of length $r_{np}$ connecting the $n$th transmitting antenna to the $p$th receiving one, and vice versa.

In (\ref{eq_link2}) only a receiving configuration with $\ell_R=\ell_T$ allows for a correct reception of the incoming OAM beam with azimuthal index $\ell_T$. In this case, expression (\ref{eq_link2}) defines, as a function of the azimuth and elevation angles of one of the two arrays, the OAM-link pattern, characterized by an on-axis maximum where the conventional radiation pattern exhibits the central null. 

Following the approach described in \cite{Cagliero}, we can evaluate the link budget for a couple of facing UCAs made of half-wave dipoles by simply inserting in (\ref{eq_link2}) the effective height formula relative to this type of antenna \cite{Orfanidis}. In the case of an $N$-element UCA, the synthesis coefficients for the generation of an OAM beam with index $\ell$ are simply fixed by requiring the phase difference between each pair of subsequent array elements to be $\delta\varphi=2\pi\ell/N$ \cite{Thide}. The relation that determines how many OAM modes such array can support is: $-N/2<\ell<N/2$. 

In order to conform our analysis to the experiment, which will be described in the following, we consider two identical 4-element facing UCAs with radius $R=1.5$ m and separated by a link distance $d=40$ m, as schematically depicted in Fig. \ref{fig_gaffo2}. Since the array elements used in the experiment are folded Yagi-Uda antennas, a simple way to address the problem consists in introducing a correction to the expression (\ref{eq_link2}) for the link budget estimation in the case of half-wave dipoles, which correspond to the driven elements of each Yagi-Uda antenna. The presence of four parasitic elements in the Yagi-Uda antennas makes the radiation pattern of the arrays more directive, focusing the emitted radiation along the direction of propagation in transmission and increasing the ability of the receiving array to intercept the incoming beam along the same direction. This effect can be taken into account by adding to the link budget an amount equal to twice the power gain difference between the Yagi-Uda antenna and a half-wave dipole. Since the considered Yagi-Uda antennas have a gain of about $7$ dBi and the gain of a half-wave dipole is estimated at $2.15$ dBi, the values obtained with expression (\ref{eq_link2}) should be approximately increased by $5+5=10$ dB.

\begin{figure}[t]
\centering
\includegraphics[trim=0 0 0 0,clip,width=3.5in]{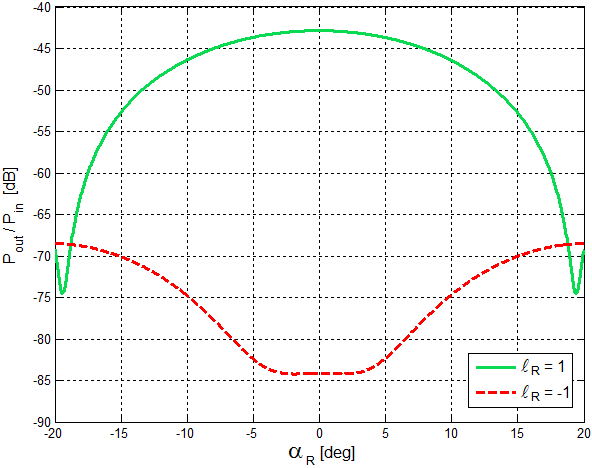}
\caption{Total received power normalized to the total radiated power {\itshape vs} the rotation angle $\alpha_R$ relative to the receiving array. The considered arrays  are 4-elements UCAs made of folded Yagi-Uda antennas and separated by a distance $d=40$ m. The OAM beam transmitted with azimuthal index $\ell_T=1$ at a frequency of $205$ MHz is received according to two different rephasing configurations.}
\label{fig_gaffo3}
\end{figure}

According to these arguments, Fig. \ref{fig_gaffo3} shows the behaviour of (\ref{eq_link2}) as a function of the angle $\alpha_R$ relative to the rotation of the receiving array around the $y$-axis passing through its center, for the couple of facing Yagi-Uda antenna arrays. Due to the orthogonality between OAM modes and having the emitted beam an azimuthal index $\ell_T=1$, only a rephasing with index $\ell_R=\ell_T=1$ maximizes the received signal at $\alpha_R=0$, while a destructive configuration occurs whenever $\ell_R\neq\ell_T$.

\begin{figure}[t]
\centering
\includegraphics[trim=0 0 0 0,clip,width=6.25in]{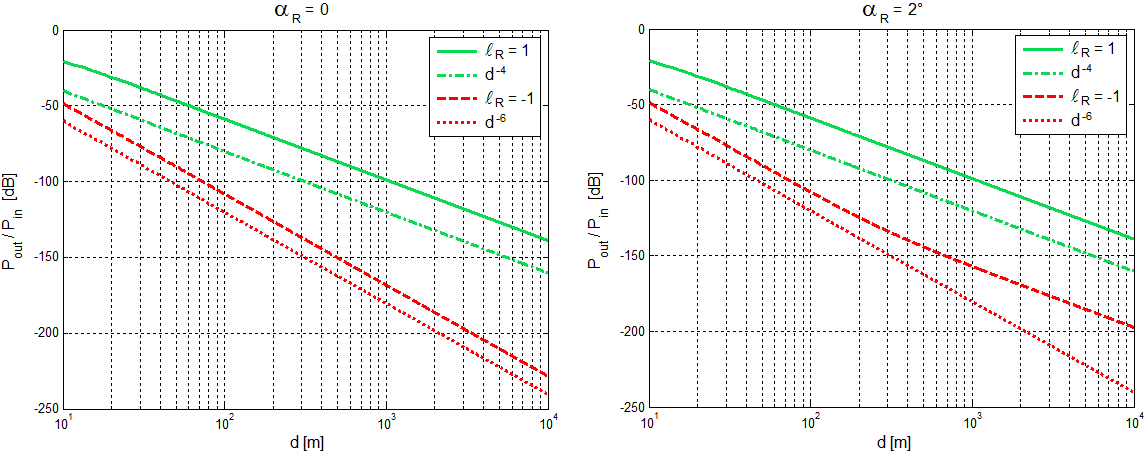}
\caption{Total received power normalized to the total radiated power {\itshape vs} the link distance $d$.}
\label{fig_gaffo4}
\end{figure}

For completeness, Fig. \ref{fig_gaffo4} reports the behaviour of formula (\ref{eq_link2}) at two fixed angles $\alpha_R$ {\itshape versus} the link distance between the Yagi-Uda antenna arrays. As can be observed, for any value of $\alpha_R$ the curve corresponding to the proper reception of the incoming OAM mode (i.e. the solid green line) follows the well-known asymptotic behaviour $d^{-2\ell_T-2}$.

For arrays of isotropic radiators, the destructive configuration $\ell_R=-\ell_T$ settles to zero at $\alpha_R=0$, where the orthogonality of the OAM modes is at its maximum, and decreases with distance according to the usual law $d^{-2\ell_T-2}$ when $\alpha_R\neq 0$. Due to the directivity of the considered array elements, this receiving configuration for the system under examination approximates the case of arrays made of isotropic elements only at great distances, as can be seen in Fig. \ref{fig_gaffo4} (dashed red line).

\section{Simulations} \label{sec_simul}

The numerical analysis performed to support the experiment was carried out via the RF module of the software {\scshape{Comsol}} Multiphysics\textsuperscript\textregistered \cite{Comsol}, which is based on the Finite Element Method (FEM) and represents a valuable tool, enabling one to take into account the system complexity as well as couplings among antennas. On the other hand, the analysis needs a complete meshing of the whole space involved in the electromagnetic transmission, so a high computational demand for large systems is required. A further problem arises from the necessity to introduce the Perfectly Matched Layer (PML), an adsorbing region which mathematically reproduces the effect of the free space outside the volume of interest where the electromagnetic transmission between the antenna arrays takes place. Being the PML absorbing efficiency numerically limited and strongly sensitive to the incidence angle, one has to enlarge the region of interest in order to prevent reflections of tangential rays on the PML walls, that would compromise the evaluation of the model parameters, especially when such quantities are small. 

We studied the electromagnetic transmission of OAM waves between two identical facing arrays of four Yagi-Uda antennas, each composed by five parallel thin metal strips and fed by a lumped port on the driven element. All the involved quantities were set according to the experimental values and the wavelength was fixed at $\lambda=1.46$ m. The receiving array orientation was parameterized in $\alpha_R$, which corresponds to the angle of rotation of the array around the $y$-axis passing through its center, while a further parameterization was provided by the distance $d$ between the transmitting and the receiving systems. 

The main goal of our simulation was to compute the OAM-link pattern as a function of the variable $\alpha_R$ in terms of a parametric sweep on the total received power, which can be expressed as a suitable combination of post-processed quantities:
\begin{equation} \label{eq_voltages}
\frac{P^{\ell_{T,R}}_{out}}{P_{in}}=\frac{\left|\frac{1}{\sqrt{N_R}}\sum^{{\scriptscriptstyle{N_R}}}\limits_{\scriptscriptstyle{p=1}} \Phi^p_{-\ell_R}V^p_{port}\right|^2}{\sum^{{\scriptscriptstyle{N_T}}}\limits_{\scriptscriptstyle{n=1}}\left|V^n_{port}\right|^2},
\end{equation}
where $\Phi^i_{\ell}$ is the $i$th synthesized coefficient for the considered OAM mode and $V^i_{port}$ represents the corresponding model port voltage, i.e. the open-circuit voltage at the terminals of the $i$th antenna. 

\begin{figure}[t]
\centering
\includegraphics[trim=0 0 0 0,clip,width=3.5in]{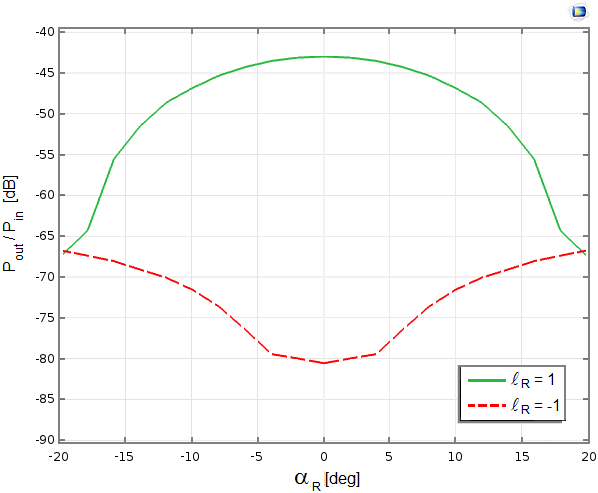}
\caption{OAM-link pattern for the transmission of an $\ell_T=1$ wave at $40$ m, extrapolated from the curve computed at $20$ m in order to cope with excessive computational demand.}
\label{fig_gaffo5}
\end{figure}

As explained before, we have to avoid possible reflections from the PML walls, a problem that is the more severe the smaller the distance between PML and antennas. Further complications lie in the fact that the OAM beam divergence angle is larger than in the $\ell=0$ case and that each reflection is accompanied by a change in sign of the azimuthal index $\ell$. For these reasons, the distance between the PML walls and the antenna arrays should be considerable, introducing a trade-off between precision in the calculated quantities and computational expense. As a matter of fact, it was verified that, in correspondence with a not enough large transmission volume, the unwanted reflections affect the curves and in particular the red one in which the incoming wave $\ell_T=1$ is read in destructive configuration $\ell_R=-1$, 
causing the disappearance of the expected central dip around $\alpha_R=0$. On the other hand, the meshing of a whole $40$ m-length transmission volume, large enough to avoid most of the reflections on the PML walls, would have required an excessive computational capacity to be tolerated by our server machine. Fortunately, beyond a given distance (which is expected to approximately match the far-field threshold) the OAM-link pattern does not show any significant changes except for a scale factor related to the beam divergence, therefore leading to the possibility of reasonable extrapolations at very high distances. The application of this useful procedure to our case is presented in Fig. \ref{fig_gaffo5}, where the OAM-link pattern relative to the transmission of an $\ell_T=1$ wave at $40$ m is extrapolated from the corresponding pattern computed at $20$ m with {\scshape{Comsol}} Multiphysics\textsuperscript\textregistered\space according to the far-field decays of the OAM beams shown in Fig. \ref{fig_gaffo4}.

\section{OAM transmission measurements}\label{sec_OAMtrans}
\subsection{Mode sorters}\label{sec_modesorters}

In order to introduce suitable phase shifts to both the transmitting and the receiving antenna arrays, a proper electronic device, called {\itshape mode sorter}, was designed with the aim of encoding/decoding up to $N-1$ OAM-modes in the case of $N$-element UCAs \cite{modesorter}.

\begin{figure}[b]
\centering
\includegraphics[trim=0 0 0 0,clip,width=6.2in]{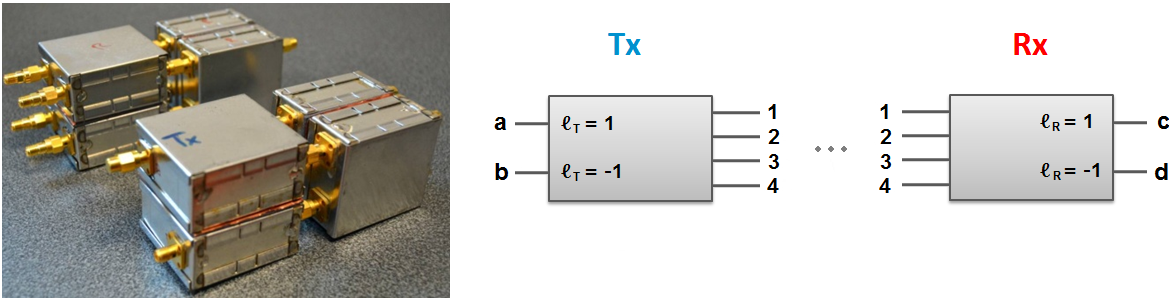}
\caption{Picture ({\it left}) and schematic illustration ({\it right}) of the mode sorters employed in the experimental work.}
\label{fig_gaffo6}
\end{figure}

Although with $N=4$ antennas the $\ell=0$ mode could be generated too, we chose to limit our investigation to just $\ell=\pm 1$ modes, for simplicity. The prototyped mode sorter, used in the experiment and schematically depicted in Fig. \ref{fig_gaffo6}, was realized in order to reproduce the following ideal behaviour: the four outputs, connected to the antennas of the transmitting UCA, determine the emission of an OAM beam with azimuthal index $\ell_T=1$ or $\ell_T=-1$, depending on whether the input port {\textbf{a}} or {\textbf{b}} is fed. For symmetry, when the considered mode sorter is used in reception, an opposite phase-weighting is introduced over the incident beam; as a result, the signals read at the output ports {\textbf{c}} and {\textbf{d}} are rephased with an index $\ell_R=1$ and $\ell_R=-1$, respectively. 

However, due to the technical limitations of the electronics, the outgoing beam cannot be intended as a pure OAM mode with index $\ell_T=\pm 1$, but rather as a linear combination of all the OAM modes supported by the considered 4-element UCA, with a prevalence of the $\ell_T=1$ or the $\ell_T=-1$ contribution depending on the input port chosen.

With the aim of realizing an experiment of a real TV signal transmission, a couple of mode sorters working on TV frequencies was implemented; the reasons of this choice lie in the availability of commercial components covering a whole TV band and in the presence of free channels in the VHF spectrum ($174\div 230$ MHz). 

To design the mode sorters circuit whatever technique can be used. Here a specially devised synthesis method that allows to avoid components with more than two ports has been employed. Most phase shifters usually set a severe limit to the bandwidth of the system; in our case with $N=4$ antennas the synthesis method enables the use of two-port $180^\circ$ and $90^\circ$ hybrids, that can be easily found in the wide-band version. The price to pay is a higher insertion loss than the one achievable by a narrow-band design.

The mode sorters were tested individually in the laboratory. The maximum amplitude unbalance among the four antenna ports was measured to be $0.2$ dB (i.e. $2\%$ vector error) and the estimated maximum phase error w.r.t. the nominal values (integer multiples of $90^\circ$) among the four antenna ports was found to be $2.8^\circ$ (i.e. $5\%$ vector error). Being an analog high frequency wide-band device with no tuning capabilities, this performance was considered as acceptable to the purpose of our experiment.
To assess an upper bound to the OAM channel insulation achievable from this couple of mode sorters, the performance of the cascade connection of the two devices was evaluated. The signal entering port {\textbf{a}} of the Tx mode sorter should be mostly conveyed to port {\textbf{c}} of the Rx mode sorter. Similarly, signal to port {\textbf{b}} should be mostly conveyed to port {\textbf{d}}. Crossed paths should experience a large attenuation.

\begin{figure}[t]
\centering
\includegraphics[trim=0 0 0 0,clip,width=3.4in]{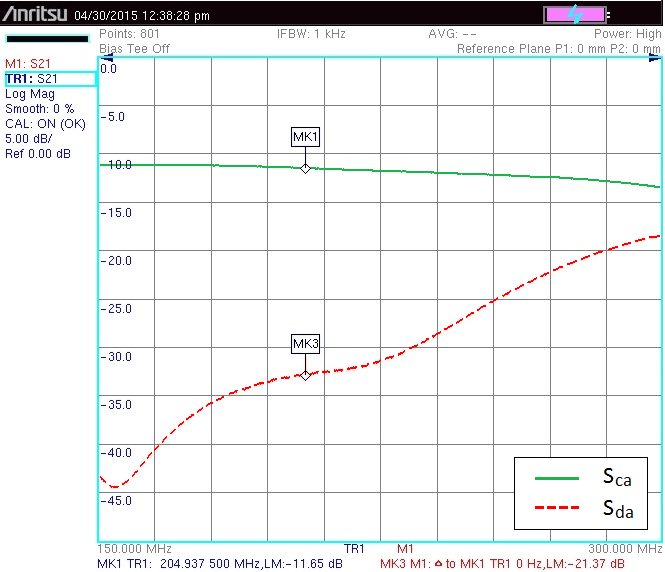}
\caption{Cascade mode sorters performance: plot of the $S_{21}$ parameter {\it vs} frequency.}
\label{fig_gaffo7}
\end{figure}

For the measurement an Anritsu MS2026C Vector Network Analyzer (VNA) was used; Fig. \ref{fig_gaffo7} shows as a function of frequency the transmission parameter ($S_{21}$) which defines, in a two-port device, the power transferred from port 1 to port 2. Scattering ($S$) parameters are widely used in microwave/RF engineering to describe the behaviour of a linear electrical network. In Fig. \ref{fig_gaffo7}, the solid green curve is the transmission parameter from {\textbf{a}} to {\textbf{c}} {\it vs} frequency, while the dashed red curve is from {\textbf{a}} to {\textbf{d}}. As can be observed, at the intended use channel frequency (markers), the channel insulation (main path/unwanted path, markers delta) is $21.4$ dB. In the over-the-air transmission, of course, this insulation will be impaired by other factors including antennas and cable unbalances, reflections on nearby objects, etc.

\subsection{The test set-up}\label{sec_testsetup}

The antenna array was chosen to be identical for both the transmitting and the receiving systems (Fig. \ref{fig_gaffo2}) and it was built using four commercial VHF-band TV reception antennas. We adopted the Fracarro BLV4F, a 4-element folded Yagi-Uda antenna, individually providing $7$ dBi gain. The array structure was built with plastic PVC pipes, obtaining a $D=3.1$ m diameter circular array. This size results in a far-field range of about $2D^2/\lambda\simeq 13$ m at the frequency of $205$ MHz. Horizontal polarization was  chosen, in order to keep feeding cables orthogonal to the dipoles of the Yagi, in the adopted mechanical arrangement.

Although less modern than the Anritsu MS2026C, a Hewlett-Packard  HP8753B Vector Network Analyzer was used to measure the RF transmission parameters; in this equipment the $S$-parameter test set section (HP85046A) can be detached and it is possible to have direct access to the vector analyzer channels. This allows to measure simultaneously the complex ratio of two inputs with respect to a reference signal $R$. The only drawback is that an external coupler must be provided to feed the reference channel $R$.

\begin{figure}[t]
\centering
\includegraphics[trim=0 0 0 0,clip,width=6.2in]{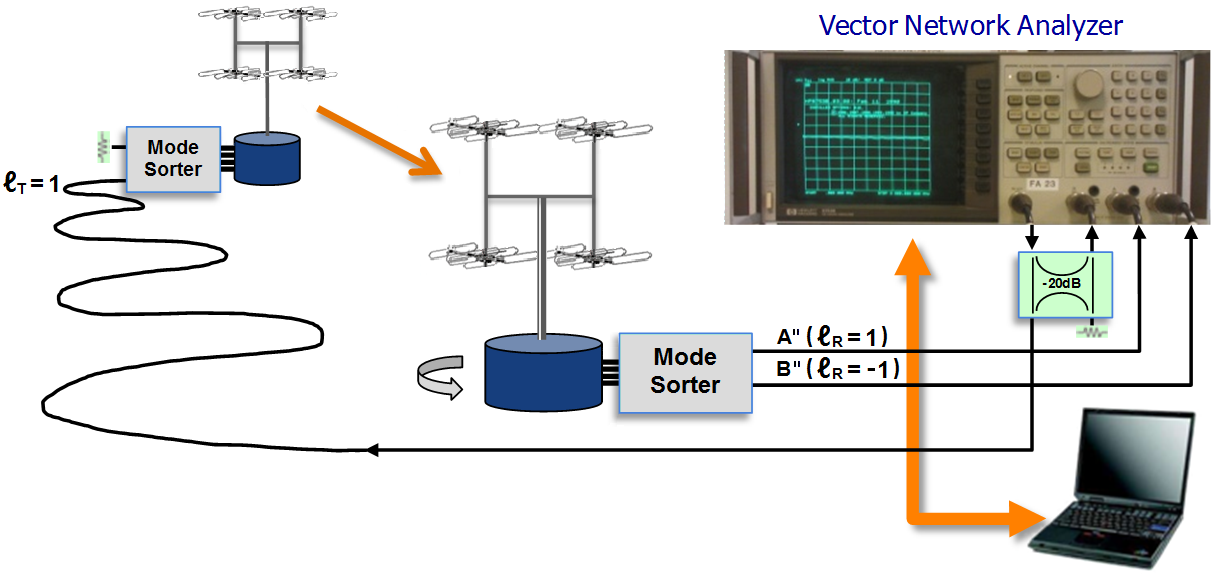}
\caption{Experiment set-up block diagram for the transmission of the $\ell_T=1$ mode.}
\label{fig_gaffo8}
\end{figure}

The VNA output is the transmitted signal source and was connected to the transmitting site via a $100$ m low loss coaxial cable, to feed one of the two ports of the Tx mode sorter. At the receiving side, the resulting signals from ports {\textbf c} ($\ell_R=1$) and {\textbf d} ($\ell_R=-1$) of the Rx mode sorter were connected to the input ports of the VNA, while the input $R$ is fed by the transmitted reference signal from the direction coupler. The set-up block diagram is shown in Fig. \ref{fig_gaffo8}.

A preliminary direct-loopback test was done, directly connecting the four outputs of the Tx mode sorter with the relevant inputs of the Rx mode sorter; the results, including the cable losses, were stored to disk for offline calibration purpose. The transmitting UCA was placed on a tripod, while the receiving one was mounted on a rotary head with $360^\circ$ goniometer. Unfortunately, the mechanic arrangement allows to explore only $\pm2$ degrees elevation. In each test session, before scanning the azimuth, the direction of the beam axis was fine tuned in both azimuth and elevation. 
In all the tests the $150\div300$ MHz frequency range was explored by the VNA. The measurement data were acquired on a PC connected to the VNA. In the further offline processing it was possible to clip the frequency range and to apply calibration and frequency average.

\subsection{The experimental tests}\label{sec_tests}

Three kinds of test were carried out. First, as a preparatory phase, the classic non-OAM (i.e. $\ell=0$) link was checked: to this purpose, both mode sorters were replaced with conventional zero-degrees, 4-way RF splitters. The calibrated measurement result, $-25.5$ dB, was found to be in good agreement with the classic link budget estimation of $-24.4$ dB (Friis formula).

Then, due to the strong sensitivity of the OAM orthogonality to misalignment, in a second step it was necessary to ensure that both the arrays were perpendicular to the line connecting their centers. For this purpose, an OAM beam with index $\ell_T=1$, characterized by a doughnut-shaped beam profile with a zero on-axis intensity, was sent to the receiving array and the voltage signals induced over all its elements were simply summed (this standard reading corresponds to the $\ell_R=0$ receiving configuration). By rigidly rotating the receiving structure, it was possible to map the intensity profile of the received beam and to determine the angular position of its minimum, which provides the proper alignment of the arrays, confirmed within a few angular degrees.

Finally, since the purpose of the third test was to study the behaviour of the two considered receiving configurations (i.e. $\ell_R=1$ and $\ell_R=-1$) when an OAM mode with index $\ell_T=1$ is transmitted, a signal $A$ provided by the VNA was introduced into the input port {\textbf{a}} of the mode sorter connected to the transmitting Yagi-Uda antennas (see Fig. \ref{fig_gaffo8}). The port {\textbf{b}} of the Tx mode sorter, corresponding to an $\ell_T=-1$ mode, was not fed and kept properly terminated.

A significant reception of the beam is expected only when the rephasing configuration is $\ell_R=\ell_T=1$. The resulting signals $A''$ and $B''$, read at the outputs of the receiving mode sorter, can be expressed in terms of the channel matrix $CH$ relative to the considered system as follows:
\begin{equation} \label{eq_f1}
\begin{pmatrix}
A'' \\
B''
\end{pmatrix}=
CH
\begin{pmatrix}
A \\
0
\end{pmatrix}\equiv
\begin{pmatrix}
\Gamma & \Delta \\
\Delta & \Gamma
\end{pmatrix}
\begin{pmatrix}
A \\
0
\end{pmatrix}, \ \ \ \ \Gamma, \Delta \in \mathbb{C}.
\end{equation}
The adopted matrix approach describes the whole process in the two-dimensional space relative to the accessible entries and outputs of the mode sorters. Moreover, since the matrix $CH$ takes into account the action of the mode sorters over the transmitted signal, it can be decomposed in the following product:
\begin{equation} \label{eq_f2}
CH=RX\cdot CH^p\cdot TX = 
\begin{pmatrix}
\alpha^\ast & \beta^\ast \\
\beta^\ast & \alpha^\ast
\end{pmatrix}^{-1}
\begin{pmatrix}
\gamma & \delta \\
\delta & \gamma
\end{pmatrix}
\begin{pmatrix}
\alpha & \beta \\
\beta & \alpha
\end{pmatrix}, \ \ \ \ \alpha,\beta,\gamma,\delta\in\mathbb{C}
\end{equation}
where the {\itshape pure} channel matrix $CH^p$ is obtained from $CH$ by removing the unwanted electronic contribution introduced by the electronics. The mode sorters were expressed in terms of the two-dimensional matrices $TX$ and $RX$, which are one the inverse conjugate of the other, due to their opposite orientation and the OAM chirality inversion at the reception. The reason for choosing bidimensional matrices lies in the possibility to express, within a tolerable margin of error, the four-dimensional transmitting outputs and receiving inputs of the mode sorters in terms of the OAM states with indices $\ell=1$ and $\ell=-1$. The validity of these assumptions, together with the bisymmetry of the mode sorter matrices, was experimentally verified on the basis of several measurements made on the devices themselves, in which known signals were sent to the various inputs, with relative acquisition of the output signals. 

\begin{figure}[t]
\centering
\includegraphics[trim=0 0 0 0,clip,width=3.5in]{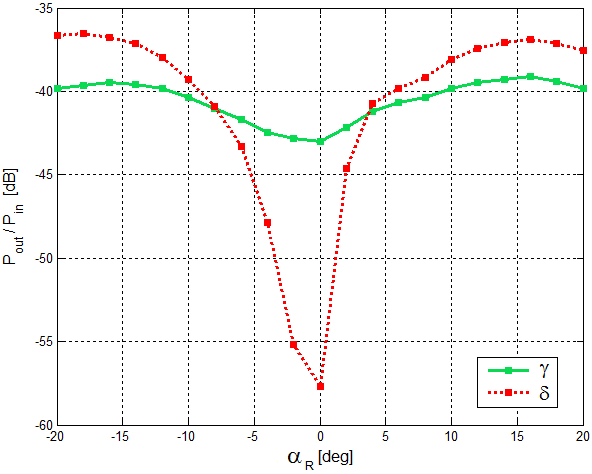}
\caption{Experimentally derived OAM-link pattern averaged over the frequency range: $200\div 210$ MHz (10 values of frequency centered around $205$ MHz).}
\label{fig_gaffo9}
\end{figure}

In (\ref{eq_f2}) the matrix elements $\gamma$ and $\delta$ provide an OAM-link budget estimation relative to the reception of the signal $A$ at the output ports of the receiving mode sorter corresponding to the $\ell_R=1$ and the $\ell_R=-1$ receiving configurations, respectively. 

During the experiment, the receiving UCA, mounted on a mobile support, was rotated azimuthally around its center by an angle $\alpha_R\in[-20^\circ,20^\circ]$, with an increment of $2^\circ$; hence, the experimental signals $A''$ and $B''$ were obtained for the different values of the rotation angle $\alpha_R$ and all these measures were repeated for $800$ frequency values in the range $150\div 300$ MHz.

\begin{figure}[t]
\centering
\includegraphics[trim=0 0 0 0,clip,width=3.5in]{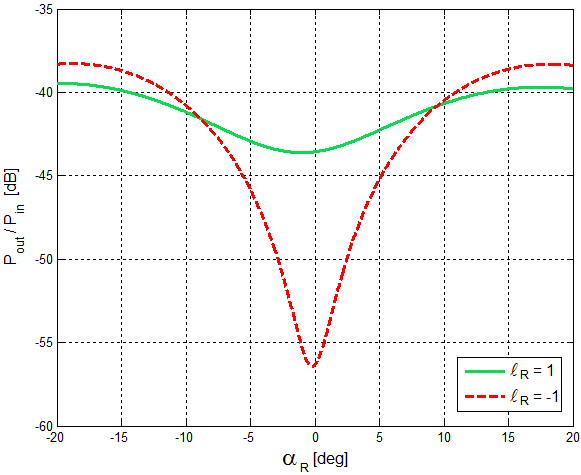}
\caption{OAM-link pattern (\ref{eq_link2}) in the presence of simulated angular inaccuracies affecting the arrays position ($\alpha_T=4^\circ$).}
\label{fig_gaffo10}
\end{figure}

In order to obtain the coefficients $\gamma$ and $\delta$ from (\ref{eq_f1}) it was necessary to analyze the action of the mode sorters alone on the transmitted signal.
For this reason, in the preliminary test, the mode sorters were directly connected one to the other, leading to the acquisition, for every value of frequency, of the signals $A'$ and $B'$, which can be expressed by:
\begin{equation} \label{eq_f3}
\begin{pmatrix}
A' \\
B'
\end{pmatrix}=
RX\cdot TX
\begin{pmatrix}
A \\
0
\end{pmatrix}=
\begin{pmatrix}
\alpha^\ast & \beta^\ast \\
\beta^\ast & \alpha^\ast
\end{pmatrix}^{-1}
\begin{pmatrix}
\alpha & \beta \\
\beta & \alpha
\end{pmatrix}
\begin{pmatrix}
A \\
0
\end{pmatrix}.
\end{equation}

Taking into account (\ref{eq_f2}), equations (\ref{eq_f1}) and (\ref{eq_f3}) allow to derive the coefficients $\gamma$ and $\delta$ for each value of angle and frequency, starting from all the experimentally known signals. 

Fig. \ref{fig_gaffo9} displays the behaviour of $\gamma$ and $\delta$, averaged over 10 frequency values relative to the $205$ MHz channel, {\itshape vs} the receiving array rotation angle $\alpha_R$. The condition of maximum phase compensation of the signal $A$ was expected at the $\ell_R=1$ output port of the receiving mode sorter when $\alpha_R=0$, in agreement with the theoretical result illustrated in Fig. \ref{fig_gaffo3}, section \ref{sec_OAMpattern}. On the contrary, Fig. \ref{fig_gaffo9} shows how the respective $\gamma$ curve presents a slight central dip; moreover, the dotted red curve $\delta$, which corresponds to the destructive receiving configuration, overcomes $\gamma$ moving away from $\alpha_R=0$ instead of settling at lower values. As we will prove below, the observed behaviour can be attributed to some experimental inaccuracies, such as unwanted small angular shifts of the arrays position with respect to their optimum reciprocal centering able to maximize the OAM orthogonality. Despite this, our experimental result shows that, at $\alpha_R=0$, the $\ell_R=1$ receiving configuration allows for a significant reception of the $\ell_T=1$ signal, while the $\ell_R=-1$ output suppresses it, ensuring an insulation of about $15$ dB.

In order to provide a theoretical confirmation of the results obtained in the experiment, formula (\ref{eq_link2}) for the link budget evaluation was considered in the presence of angular shifts affecting the optimum pointing of the {\it transmitting} array. Fig. \ref{fig_gaffo10} shows the behaviour of the OAM-link pattern, given by expression (\ref{eq_link2}), as a function of the rotation angle $\alpha_R$, when the transmitting array is azimuthally rotated around the $y$-axis passing through its center by an angle $\alpha_T=4^\circ$. As can be observed, taking into account a possible pointing error, the resulting link budget plot (Fig. \ref{fig_gaffo10}) becomes very similar to the experimental one (see Fig. \ref{fig_gaffo9}). Although the receiving array was mounted on a goniometer characterized by an angular sensitivity of about $1^\circ$, the mechanical arrangement of the transmitting array did not ensure a sufficient accuracy to avoid errors of the order of a few degrees; therefore we decided not to repeat the experiment.

\begin{table}[t]
\caption{\label{ex}Link budget estimated with four methods for the considered couple of facing UCAs composed by four folded Yagi-Uda antennas. The receiving configurations are $\ell_R=\ell_T\equiv\ell=0,1$.}
\label{tab_linkbudget}
\begin{center}
\begin{tabular}{lll}
\br
Link budget (dB) & $\ell=0$ & $\ell=1$ \\
\mr
Theoretical approach         &  -24.54      &  -42.85     \\
Friis equation               &  -24.40      &  $\ \ \ $ - \\
{\scshape Comsol} simulation &  -24.69      &  -42.96     \\
Experiment                   &  -25.50      &  -42.99     \\
\br
\end{tabular}
\end{center}
\end{table}

Table \ref{tab_linkbudget} reports the link budget evaluated with the different approaches considered in our analysis, in correspondence of an optimal antenna pointing. 
A good agreement among the theoretical, numerical and experimental results is achieved.

\section{Experimental transmission of two digital TV signals} \label{sec_experiment}

As well known, the crucial factor in a RF link performance is the ratio $C/(N+I)$ where $C$ is the received signal power, $N$ is the receiver noise power, and $I$ is any incoherent disturbance in the signal band. In our case, of course, the leakage from the other OAM mode can be considered incoherent (noise-like), since the two modulating data streams are statistically independent. For a short link, $N$ is many orders of magnitude lower than $I$, then $C/(N+I)\simeq C/I$, the latter being the OAM insulation, hereinafter referred to as Signal-to-Noise Ratio (SNR).

\begin{table}[b]
\caption{Relevant bit rates and SNR required for QEF reception reported here for some different modulation and coding schemes.}
\label{tab_bitrate}
\begin{center}
\begin{tabular}{ccccc}
\br
{Modulation} & 
{Code rate}  &
{Required SNR for} &
{Bitrate (Mbit/s)} &
{Bitrate (Mbit/s)} \\

& 
&
{QEF operation} &
{$@\:\Delta/TU=1/32$,} &
{$@\:\Delta/TU=1/32$,} \\

&
&
&
{$7$ MHz VHF mode} &
{$8$ MHz UHF mode} \\

\mr
QPSK         &  7/8  &  7.7  &  9.237  &  10.56  \\
16-QAM       &  1/2  &  8.8  &  10.556 &  12.06  \\
16-QAM       &  2/3  &  11.1 &  14.075 &  16.09  \\
16-QAM       &  3/4  &  12.5 &  15.834 &  18.10  \\
16-QAM       &  5/6  &  13.5 &  17.594 &  20.11  \\
16-QAM       &  7/8  &  13.9 &  18.473 &  21.11  \\
64-QAM       &  1/2  &  14.4 &  15.834 &  18.10  \\
\br
\end{tabular}
\end{center}
\end{table}

The  DVB-T standard \cite{DVBT} allows the network operator to tailor the modulation scheme and error correction code rate to the available SNR. In Table \ref{tab_bitrate} a few configurations extracted from \cite{DVBT} are reported, with the relevant bit rates and SNR required for ``Quasi Error Free'' (QEF) \footnote{Quasi Error Free (QEF) means less than one uncorrected error event per hour, corresponding to a BER (Bit Error Rate) of $10^{-11}$.} reception.

\begin{figure}[t]
\centering
\includegraphics[trim=0 0 0 0,clip,width=5.35in]{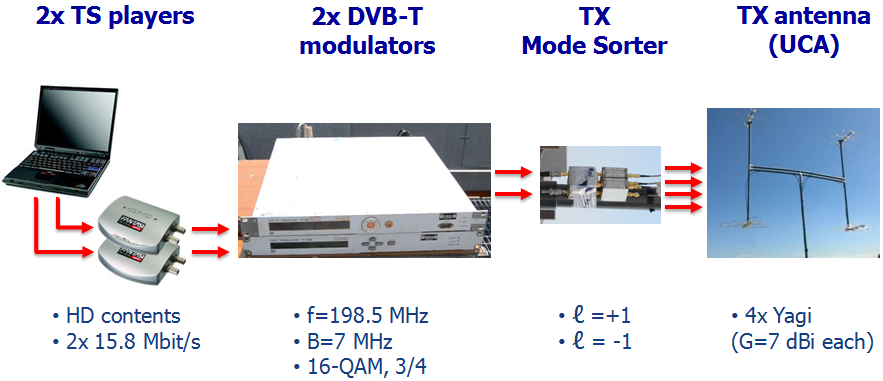}
\caption{Diagram of the transmitting installation.}
\label{fig_gaffo11}
\end{figure}

Actually, such declared performance is simulated, hence an extra implementation loss (roughly $1$ dB) is encountered in commercial TV receivers. In order to allow some margin to the expected $15$ dB value, the modulation 16-QAM with code rate $3/4$ was adopted. This allowed to transmit two $15.8$ Mbit/s audio/video streams. In Fig. \ref{fig_gaffo11} the structure of the transmitting installation of the experiment is depicted and the main parameters are summarized.

\begin{figure}[b]
\centering
\includegraphics[trim=0 0 0 0,clip,width=4.7in]{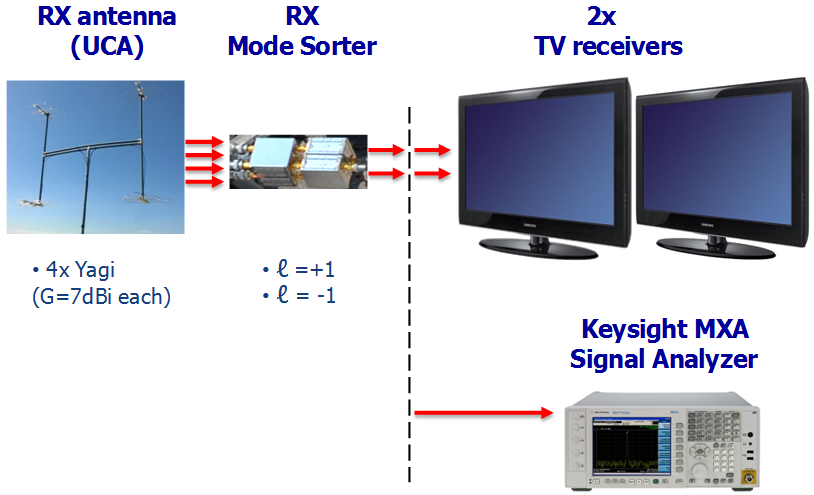}
\caption{Diagram of the receiving installation.}
\label{fig_gaffo12}
\end{figure}

The two different TV streams, $15.8$ Mbit/s each, with high-definition (HD) audio/video programmes, previously stored on a PC, were played out simultaneously by means of two USB to ASI (Asynchronous Serial Interface, \cite{ASI}) adapters (DekTec, DTU-245). The ASI streams were modulated by two PT5780 DVB-T modulators, both operating on nominal $198.5$ MHz, at $7$ MHz channel bandwidth. The RF output power was $1$ mW each. 

The diagram of the receiving installation is shown in Fig. \ref{fig_gaffo12}. The signals received by the antenna system are separated in $\ell=1$ and $\ell=-1$ OAM  components by the mode sorter, the latter feeding two identical commercial DVB-T receivers. Both TV sets were tuned to $198.5$ MHz channel. After the reception experiment, each TV receiver was disconnected and the signal was routed to a Keysight MXA (N9020A) signal analyzer with DVB-T/T2 analysis option to evaluate the received signal quality.

\section{Experiment results and discussion} \label{sec_expRes}

\begin{figure}[t]
\centering
\includegraphics[trim=0 0 0 0,clip,width=3.8in]{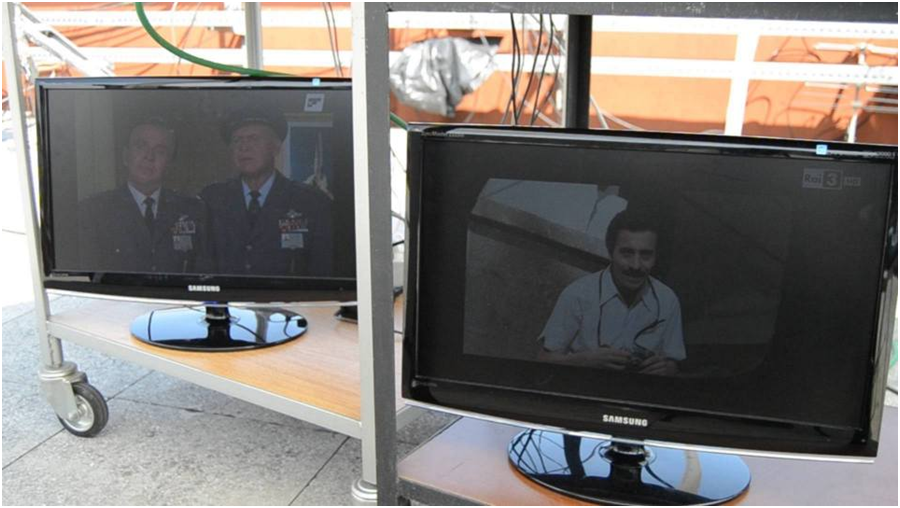}
\caption{A picture of the two TV channels simultaneously received.}
\label{fig_gaffo13}
\end{figure}

Assuming a correct antenna pointing, both TV receivers proved to simultaneously receive the relevant TV programmes (Fig. \ref{fig_gaffo13}). However, the experiment was carried out in a quite windy day and the limited mechanic rigidity of the PVC antenna mast did not help to keep optimal and constant aiming. 

\begin{figure}[b]
\centering
\includegraphics[trim=0 0 0 0,clip,width=5in]{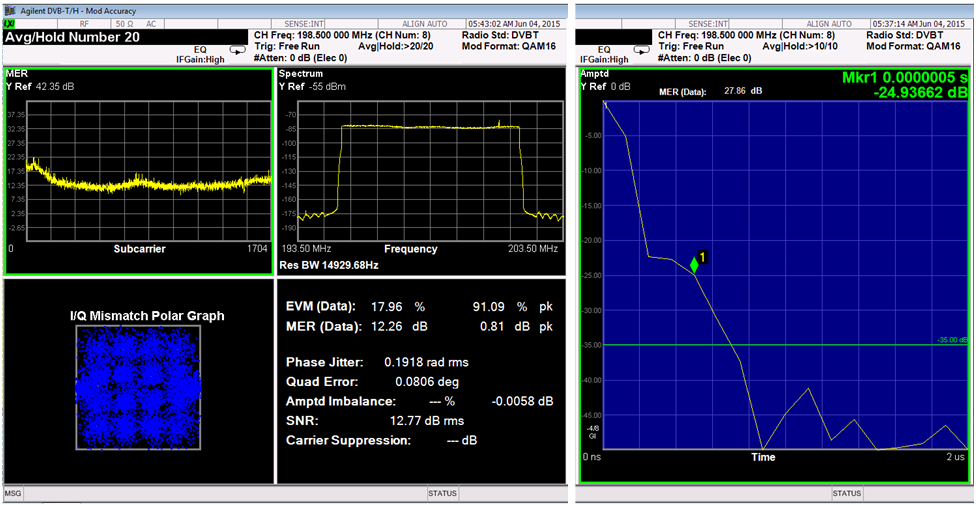}
\caption{{\it Left}: a snapshot of the MXA signal analyzer is reported, in a moment when the measured SNR was $12.77$ dB. {\it Right}: Channel Impulse Response (CIR) for the $\ell=1$ OAM channel.}
\label{fig_gaffo14}
\end{figure}

In presence of noise and/or interference, the digital TV does not exhibit errors on the screen until the SNR gets very close to a critical threshold. Hence, the margin in received signal quality cannot be estimated from the screen image, but must be evaluated only by instrumental means. The MXA signal analyzer with DVB-T/T2 option is able to measure several quality parameters of the received signal. In particular, this instrument can estimate the SNR and the MER \footnote{The \emph{Modulation Error Ratio}, MER, is the ratio of the root mean square (RMS) power of the reference vector to the power of the error vector. It is defined in decibels as: MER $= 10\log_{10}(P_{signal}/P_{error})$. SNR and MER have similar meaning, but MER is estimated on the vector space of signals.}. In Fig. \ref{fig_gaffo14} the measured Channel Impulse Response (CIR) for the $\ell=1$ OAM channel is shown. The main peak, Time = $0$, Amptd = $0$ dB, is related to direct propagation path; at about $300$ ns other components are seen at $-22$ dB level. This indicates that some propagation paths with about $300$ ns delay were present, $22$ dB below the main signal level (for $\ell=1$) and gives  an idea of the reflecting environment around the experimental site: $300$ ns time delay indicates a $100$ m path distance difference. 

During our observation with MXA, the insulation between the OAM channels fluctuated roughly from $11.2$ dB (poor aiming) to $15$ dB (good aiming) due to the wind. The latter value is in line with the insulation found during the instrumental tests. In correspondence of strong wind episodes some error could be seen on the TV screen. In that case, the problem is solved switching to the 16-QAM $2/3$ modulation. Moreover, it is important to emphasize that a device with adaptive interference cancellation capability would be able to reduce the inaccuracies caused both by a non-perfect aiming and by the presence of wind, improving the channels insulation.

\section{Conclusions} \label{sec_concl}
A numerical and experimental study concerning OAM-based transmissions between uniform circular arrays of Yagi-Uda antennas has been presented. The novelties of this work lie in the possibility 1) to implement a multimode communication at the same frequency and in the far-field region using a {\it single} antenna system (UCA) at both the transmitting and the receiving site; 2) to perform an efficient transmission considering a minimal number of array elements; 3) to validate the {\it OAM-link pattern} concept both numerically and experimentally and, lastly, 4) to test the functionality of the mode sorters as OAM multiplexing and demultiplexing networks in antenna arrays, showing a satisfactory mode insulation and wide-band operation.

While bearing in mind that the well-known power decay of OAM beams with respect to the link distance $d$ (i.e. $d^{-2\ell-2}$) represents an unavoidable problem, the main goal of this article was to show the feasibility of the transmission/reception of two digital television (DVB-T) signals encoded as OAM $\ell=1$ and $\ell=-1$ modes at the same frequency, over a link distance longer than three times the far-field range.

Finally, it should be noted that the achieved OAM mode insulation is acceptable for the purposes of our experiment, but a dedicated post-processing cancellation technique would greatly improve the performance in a more general application scenario.

\ack
We wish to thank Prof. Paolo Gambino and Prof. Roberto Tateo, from the University of Torino, Department of Physics, for the preliminary discussions and the colleagues Giovanni Bongiovanni and Franco Casalegno of the Centre for Research and Technological Innovation, RAI Radiotelevisione Italiana, for the help provided in the measurements during the experimental phase of the work.

\end{document}